\documentclass[12pt]{iopart}

\usepackage{graphicx}  % figures
\usepackage{amssymb}   % for \mathbb{R}
\graphicspath{{./}{./figures/}}

\begin{document}

\title[Generating functionals for latching dynamics]{Generating 
functionals for autonomous latching 
dynamics in attractor relict networks}

\author{Mathias Linkerhand, Claudius Gros}

\address{Institute for Theoretical Physics,
Goethe University Frankfurt, Germany}
\ead{gros07[at]itp.uni-frankfurt.de}

\begin{abstract}
Well characterized sequences of dynamical states play an
important role for motor control and associative 
neural computation in the brain. Autonomous dynamics 
involving sequences of transiently stable states
have been termed associative latching in the context
of grammar generation. We propose that generating functionals
allow for a systematic construction of dynamical networks
with well characterized dynamical behavior, such as
regular or intermittent bursting latching dynamics.

Coupling local, slowly adapting variables to an attractor
network allows to destabilize all attractors, turning them
into attractor ruins. The resulting attractor relict network
may show ongoing autonomous latching dynamics. We propose to use
two generating functionals for the construction of attractor 
relict networks. The first functional is a simple Hopfield 
energy functional, known to generate a neural attractor
network. The second generating functional, which we denote
polyhomeostatic optimization, is based on 
information-theoretical principles, encoding the information 
content of the neural firing statistics. Polyhomeostatic
optimization destabilizes the attractors of the Hopfield
network inducing latching dynamics.

We investigate the influence of stress, in terms of conflicting 
optimization targets, on the resulting dynamics. Objective 
function stress is absent when the target level for the mean 
of neural activities is identical for the two generating 
functionals and the resulting latching dynamics is then found 
to be regular. Objective function stress is present when the 
respective target activity levels differ, inducing intermittent 
bursting latching dynamics. We propose that generating functionals
may be useful quite generally for the controlled 
construction of complex dynamical systems.

\end{abstract}

\maketitle

%%%%%%%%%%%%%%%%%%%%%%%%%%%%%%%%%%%%%%%%%%%%%%%%%%%%%%%%%%%5
%%%%%%%%%%%%%%%%%%%%%%%%%%%%%%%%%%%%%%%%%%%%%%%%%%%%%%%%%%%5
\section{Introduction}

The use of objective functions for the formulation of 
complex systems has seen a study surge of interest.
Objective functions, in particular objective functions
based on information theoretical principles
\cite{ay2008,ay2012,sporns2006,bell1995,becker1996,triesch05a,markovic10},
are used increasingly as generating functionals for the 
construction of complex dynamical and cognitive systems.
There is then no need to formulate by hand equations of 
motion, just as it is possible, in analogy, to generate 
in classical mechanics Newton's equation of motion from 
an appropriate Lagrange function. When studying dynamical 
systems generated from objective functions encoding general 
principles, one may expect to obtain a deeper understanding
of the resulting behavior. The kind of generating functional
employed also serves, in addition, to characterize the
class of dynamical systems for which the results obtained may 
be generically valid.

Here we study the interplay between two generating functionals.
The first generating functional is a simple energy functional.
Minimizing this objective function one generates a neural
network with predefined point attractors, the Hopfield net
\cite{hopfield1982,hopfield1984}. The second generating functional
describes the information content of the individual neural
firing rates. Minimizing this functional results in
maximizing the information entropy \cite{triesch05a,markovic10}
and in the generation of  adaption rules for the intrinsic neural
parameters, the threshold and the gain. This principle has 
been denoted polyhomeostatic optimization \cite{markovic10,markovic12}, 
as it involves the optimization of an entire
function, the distribution function of the time-averaged
neural activities. 

We show that polyhomeostatic optimization destabilizes all
attractors of the Hopfield net, turning them into attractor
ruins. The resulting dynamical network is an attractor relict
network and the dynamics involves sequences of continuously 
latching transient states. This dynamical state is characterized
by trajectories slowing down close to the succession of attractor
ruins visited consecutively. The two generating functionals can
have incompatible objectives. Each generating functional, on its
own, would lead to dynamical states with certain average
levels of activity. Stress is induced when these two target 
mean activity levels differ. We find that the system responds
to objective function stress by resorting to intermittent bursting,
with laminar flow interseeded by burst of transient state
latching.

Dynamical systems functionally equivalent to attractor
relict networks have been used widely to formulate dynamics
involving on-going sequences of transient states. Latching 
dynamics has been studied in the context of the grammar 
generation with infinite recursion \cite{russo2008,akrami2012}
and in the context of reliable sequence generation 
\cite{horn1989,sompolinsky1986}. Transient state latching
has also been observed in the brain \cite{abeles1995,ringach2003} 
and may constitute an important component of the internal
brain dynamics. This internal brain dynamics is autonomous,
ongoing being modulated, but not driven, by the sensory input
\cite{fiser2004,maclean2005}. In this context an attractor
relict network has been used to model autonomous neural
dynamics in terms of sequences of alternating neural firing 
patterns \cite{gros2007}. The modulation of this type of internal 
latching dynamics by a stream of sensory inputs results,
via unsupervised learning, in a mapping of objects present in
the sensory input stream to the preexisting attractor ruins
of the attractor relict network \cite{gros09,gros10}, the
associative latching dynamics thus acquiring semantic content.

%%%%%%%%%%%%%%%%%%%%%%%%%%%%%%%%%%%%%%%%%%%%%%%%%%%%%%%%%%%5
%%%%%%%%%%%%%%%%%%%%%%%%%%%%%%%%%%%%%%%%%%%%%%%%%%%%%%%%%%%5
\section{Generating functionals}

We consider here $N$ rate encoding neurons in continuous time,
with a non-linear transfer function $g(x)$,
\begin{equation}
y_i \ = \ g_i(x_i) \ = \ \frac{1}{1+e^{a_i (b_i-x_i)}}~,
\label{transfer_function}
\end{equation}
where $x_i(t)\in {\cal R}$ is the membrane potential
and $y_i(t)\in[0,1]$ the firing rate of the $i$th
neuron. The dynamics of the neural activity is 
\begin{equation}
\label{xdot}
\dot x_i \ =\  -\Gamma x_i + \sum_{j=1}^{N} w_{i j} y_j~,
\end{equation}
which describes a leaky integrator, with $\Gamma$ being 
the leak rate and the $w_{i j}$ the inter-neural synaptic 
weights. This kind of dynamics can be derived minimizing 
a simple energy functional\footnote[1]{Strictly 
speaking, minimizing (\ref{energy_functional}), one obtains
$
\dot x_i  =  -\Gamma x_i + a_i(1-y_i)y_i\sum_{j} w_{i j} y_j
$ and not (\ref{xdot}). It is however custom to neglect the
extra term $a_i(1-y_i)y_i$. The term $\propto\Gamma$ in 
(\ref{energy_functional}) restricts runaway growth of the 
membrane potentials.}
\begin{equation}
-\frac{1}{2}\sum_{ij}\big(y_i w_{ij}y_j\big)^2+\frac{\Gamma}{2}\sum_i x_i^2
\label{energy_functional}
\end{equation}
with respect to the membrane potential $x_i$ \cite{hopfield1982,hopfield1984}.
When using an energy functional to derive (\ref{xdot}), the
resulting synaptic weights are necessarily symmetrized, viz
$w_{ij}=w_{ji}$. Here we do indeed consider only symmetric
synaptic links, the dynamics of a network of leaky integrators
would however remain well behaved when generalizing to 
non-symmetric weight matrices.

The transfer function (\ref{transfer_function}) contains 
two intrinsic parameters, $a$ and $b$, which can be either
set by hand or autonomously adapted, an approach considered 
here. The intrinsic parameters determine the shape of the 
individual firing-rated distributions $p_i(y)$, given
generically by
\begin{equation}
p_i(y) \ =\ \frac{1}{T}\int_0^T \delta\big(y-y_i(t-\tau)\big)
\,\mathrm{d}\tau~,
\label{def_p}
\end{equation}
where $\delta(\cdot)$ is the Dirac $\delta$-function,
and $T\to\infty$ the observation time-interval.
The Kullback-Leibler divergence \cite{grosBook}
\begin{equation}
D_{\mathrm{KL}}(p_i, q) \ =\  \int_0^{1} \mathrm{d}y\,p_i(y) 
\log \frac{p_i(y)}{q(y)} 
\label{KL}
\end{equation}
measures the distance between the neural firing rate
distribution $p_i(y)$ and a normalized distribution $q(y)$,
with $D_{\mathrm{KL}}\ge0$ generically and 
$D_{\mathrm{KL}}=0$ only for $p_i(y)=q(y)$.
Gaussian distributions
\begin{equation}
q(y)\ \propto\ \exp(\lambda_1 y + \lambda_2 y^{2}) 
\label{q_y}
\end{equation}
maximize the information entropy whenever mean and
standard deviation are given \cite{grosBook}.
Minimizing the Kullback-Leibler divergence (\ref{KL})
with respect to the individual intrinsic parameters
$a_i$ and $b_i$ is hence equivalent to optimizing
the information content of the neural activity
\cite{triesch05a}. Using variational calculus 
one obtains \cite{markovic10,markovic12,linkerhand2012}
\begin{equation}
\begin{array}{rcl}
\dot a_i & =&  \epsilon_a \Big( {1}/{a_i} + (x_i-b_i) 
      \big[ 1 - 2 y_i + \big( \lambda _1 + 2 \lambda _2 y_i \big)
       \left( 1 - y_i \right) y_i \big] \Big)_{\phantom{|}} \\
\dot b_i &=& \epsilon _b \Big( - a_i\, \big[ 1 - 2 y_i + 
\left( \lambda _1 + 2 \lambda _2 y_i \right) \left( 1 - y_i \right) y_i \big] 
\Big)~,
\label{abdot}
\end{array}
\end{equation}
where the $\epsilon_a$ and $\epsilon_b$ are adaption rates.
When these adaption rates are small, with respect to the
time scale of the cognitive dynamics (\ref{xdot}),
an implicit time average is performed
\cite{markovic12,linkerhand2012}, compare (\ref{def_p}),
and the respective firing rate distribution $p_i(y)$ 
approaches the target distribution function $q(y)$.
The adaption rules (\ref{abdot}) implement the optimization 
of an entire distribution function, and not of a single
scalar quantity, and are hence equivalent to a polyhomeostatic
optimization process \cite{markovic10}.

The original network (\ref{xdot}) has point attractors, as
given by $\dot x_i=0,\ \forall i$. These attractors
are destabilized for any $\epsilon_a,\epsilon_b>0$, the
coupled system (\ref{xdot}) and (\ref{abdot}) has
no stationary points, as defined by $\dot x=0=\dot a=\dot b$.
Stationary point attractors lead to neural firing statistics
far from the target distribution (\ref{q_y}) and to a large 
Kullback-Leibler divergence $D_{\mathrm{KL}}$, and hence
to finite stochastic gradients (\ref{abdot}). The principle
of polyhomeostatic adaption is hence intrinsically destabilizing.

In deriving the stochastic adaption rules (\ref{abdot}) one
rewrites the Kullback-Leibler divergence $D_{\mathrm{KL}}$
as an integral over the distribution $p_i(x)$ of the respective
membrane potential $x_i$, namely as
$D_{\mathrm{KL}}=\int \mathrm{d}x\,p_i(x)\,d(x)$, with an appropriate
kernel $d(x)$. The optimal overall adaption rules depend on 
the specific shape of $p_i(x)$. The optimal adaption rates not
dependent on the distribution of the membrane potential
are, on the other side, given by minimizing the kernel,
$\dot a\propto\partial d(x)/\partial a$ and
$\dot b\propto\partial d(x)/\partial b$ respectively,
which leads to the adaption rates (\ref{abdot}), which are 
instantaneous in time \cite{markovic10,markovic12,linkerhand2012}.

%%%%%%%%%%%%%%%%%%%%%%%%%%%%%%%%%%%%%%%%%%%%%
\begin{figure}[t]
\centering
\noindent
%\raise -4.0cm \hbox{
\raise 1.5cm \hbox{
\includegraphics[width=0.25\columnwidth]{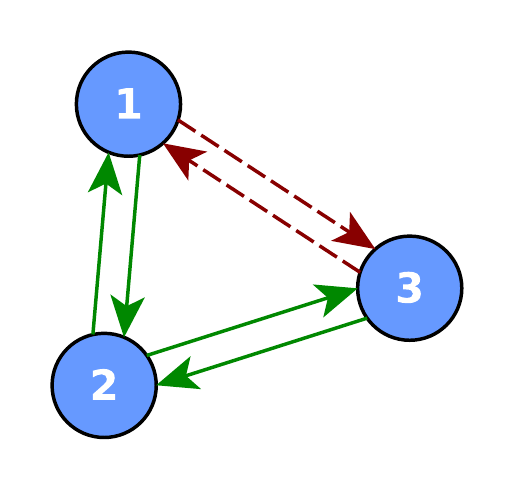}
                 }
\hspace{-1ex}
\includegraphics[width=0.60\columnwidth,angle=0]{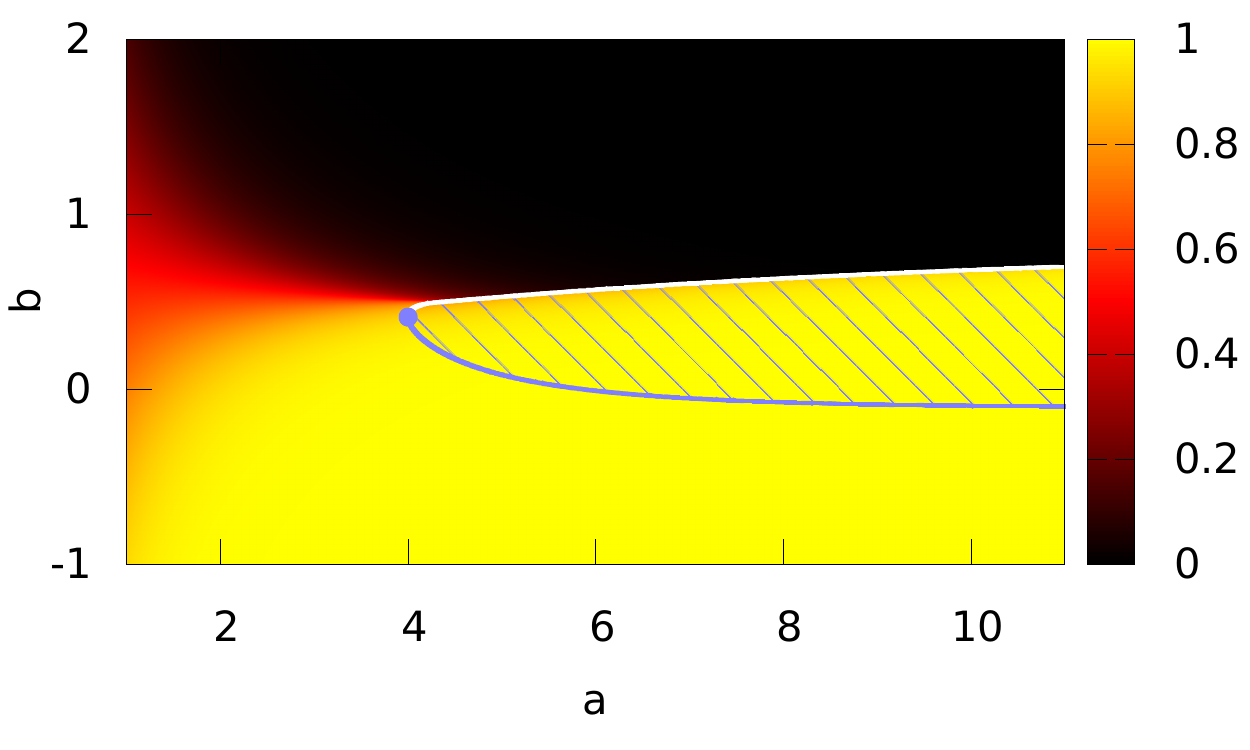}
\vspace{-1cm}
\caption{Left: A three-site graph with symmetric
excitatory (solid green lines) and inhibitory connections 
(red dashed lines). 
Right: The corresponding phase diagram, for
fixed $\Gamma = 1$ and $w^+=1=w^-$, as function of
the gain $a=a_i$ and threshold $b=b_i$, for $i=1,2,3$.
For large/small thresholds the sites tend to be inactive/active.
The color encodes the activity of the most active neuron. 
Inside the shaded area there are two stable attractors, 
outside a single, globally attracting state. The binary
representations of the two non-trivial attractors are (1,1,0) 
and (0,1,1), where 1/0 denotes an active/inactive site.
The lower and upper lines (blue/white) of the phase transition
are of second and first order respectively, with the
lower phase transition line determiend by 
$1=a\tilde y(1-\tilde y)$ where $\tilde y\equiv y_1=y_3$ 
is the activity of the single fixpoint at site one and three. 
The two critcal lines meet at $(a_c,b_c)=(4,0.413)$
(blue dot).
        }
\label{fig:3-site_phaseDiagram}
\end{figure}
%%%%%%%%%%%%%%%%%%%%%%%%%%%%%%%%%%%%%%%%%%%%%

%%%%%%%%%%%%%%%%%%%%%%%%%%%%%%%%%%%%%%%%%%%%%%%%%%%%%%%%%%%5
%%%%%%%%%%%%%%%%%%%%%%%%%%%%%%%%%%%%%%%%%%%%%%%%%%%%%%%%%%%5
\section{Attractor relicts in a three site network}

Starting, we consider now one of the simplest
possible networks having two distinct attractors
in terms of minima of the energy functional
(\ref{energy_functional}).
The dynamics of the 3-site network illustrated in 
Fig.~\ref{fig:3-site_phaseDiagram} is given by
\begin{equation}
\begin{array}{rcl}
\dot{x}_1 &=& -\Gamma x_1 + w^{+} y_2 - w^{-} y_3 \\
\dot{x}_2 &=& -\Gamma x_2 + w^{+} \left( y_1 + y_3 \right) \\
\dot{x}_3 &=& -\Gamma x_3 + w^{+} y_2 - w^{-} y_1 
\end{array}\qquad\quad
w = \left(\begin{array}{ccc}
0 & w^{+} & -w^{-} \\ 
w^{+} & 0 & w^{+} \\ 
-w^{-} & w^{+} & 0 
\end{array}\right)~,
\label{three_sites}
\end{equation}
with $w^{+} > 0$ and $w^{-} > 0$ denoting the excitatory 
and inhibitory link strength respectively. For fixed intrinsic 
parameters $a$ and $b$, viz for vanishing adaption rates
$\epsilon_a,\epsilon_b=0$, the network has two possible
phases, as shown in Fig.~\ref{fig:3-site_phaseDiagram}.
There is either a single global attractor, with
activities $y_i$ determined mostly by the value of the
threshold $b$, or two stable attractors, with either 
$y_1$ or $y_3$ being small and the other two firing rates large.
We will discuss the phase diagram in detail in
Sect.~\ref{subsec:phase-baundary}.

%%%%%%%%%%%%%%%%%%%%%%%%%%%%%%%%%%%%%%%%%%%%%
\begin{figure}[t]
\centering
\includegraphics[width=0.8\columnwidth]{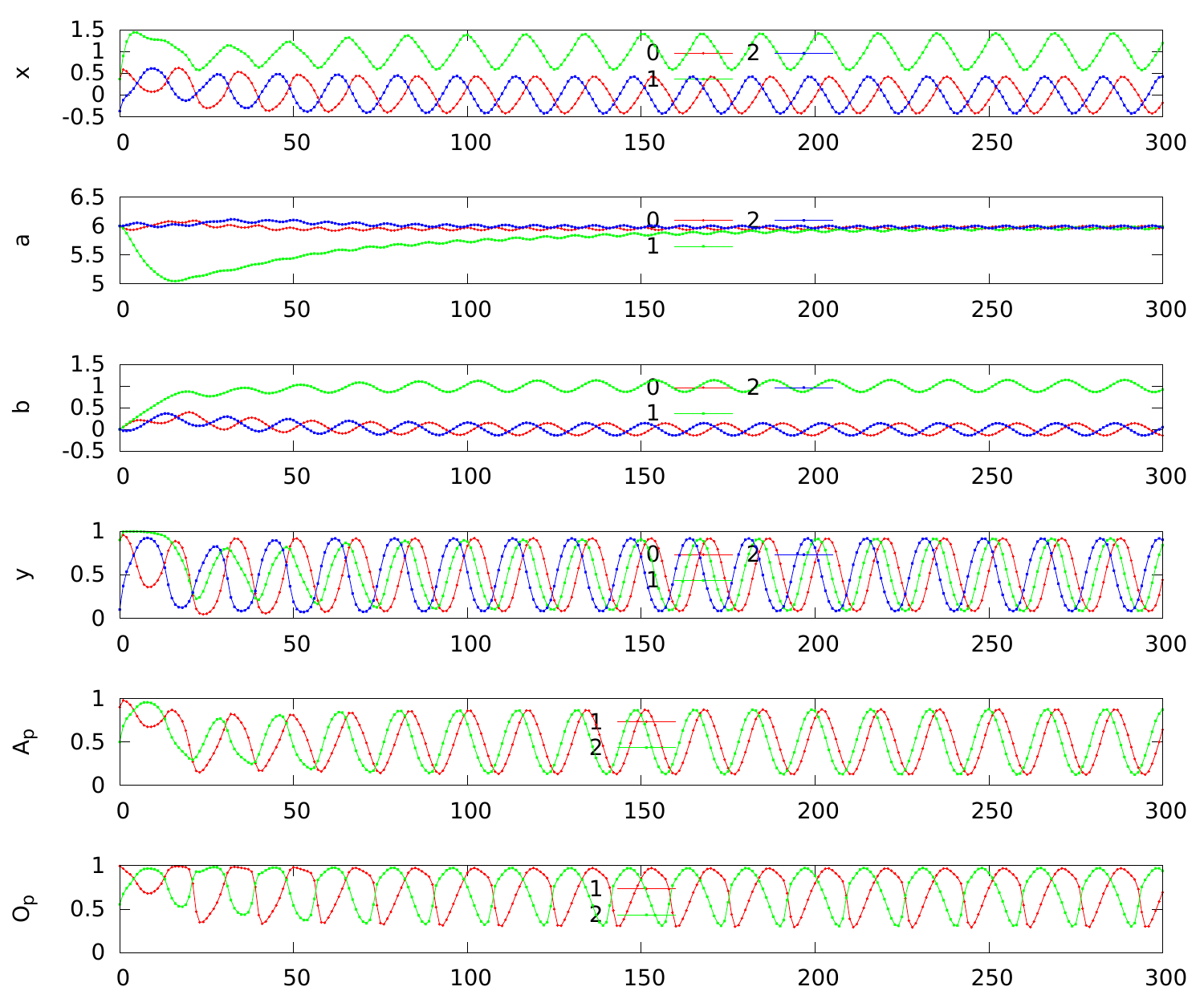}
\caption{Time series of the three-site network shown in
Fig.~\ref{fig:3-site_phaseDiagram}, for $\Gamma = 1$,
$\epsilon_a = 0.1$, $\epsilon_b = 0.01$, $\lambda_1 = 0$, 
$\lambda_2 = 0$ and $w^{\pm}=1$.
The dynamics is given by (\ref{xdot}) and (\ref{abdot}).
The dynamics retraces periodically the original attractor states,
as one can see from the oscillation of the 
overlap $O_p$ and the $A_p$, compare Eqs.\ (\ref{O_p}) and
(\ref{A_p}), between the patterns of neural and attractor
activities.
}
\label{fig:three_site_neurons_time_series}
\end{figure}
%%%%%%%%%%%%%%%%%%%%%%%%%%%%%%%%%%%%%%%%%%%%%

The point attractors present for vanishing intrinsic adaption 
become unstable for finite adaption rates 
$\epsilon_a,\epsilon_b>0$. Point attractors, defined by 
$\dot x_i=0=\dot y_i$, have time-independent neural 
activities. The objective (\ref{KL}), to attain a
minimal Kullback-Leibler divergence, can however 
not be achieved for constant firing rates $y_i$. The 
polyhomeostatic adaption (\ref{abdot}) hence forces the 
system to become autonomously active, as seen
in Fig.~\ref{fig:three_site_neurons_time_series}.
We denote with 
\begin{equation}
\xi^p \ =\ (x_1^p,\dots,x_N^p),
\qquad\quad
||\xi^p|| = \sqrt{\sum_{i=1}^N \left(\xi_i^p\right)^2}
\label{xi_def}
\end{equation}
the binary activity patterns of the attractors 
for the $N=3$ network, $\xi^{1}=(1,1,0)$ and
$\xi^{2}=(0,1,1)$. In the following we want to investigate to
which degree the original binary attractors are retraced. For
this propose we define two criteria, the first being the
overlap $O_p\in[0,1]$,
\begin{equation}
 O_p \ = \ \frac{\left\langle \xi^p, y \right\rangle }
{\left\| \xi^p \right\| \left\| y \right\| } ~,
\qquad\quad
\left\langle \xi^p, y \right\rangle = \sum_{i=1}^N \xi_i^p y_i~,
\label{O_p}
\end{equation}
which corresponds to the cosine of the angle between the
actual neural activity $(y_1,\dots,y_N)$ and the
attractor state $\xi^p$. As a second measure, of
how close the actual activity pattern and the original
attractors are, we consider the reweighted 
scalar product $A_p\in[0,1]$ 
\begin{equation}
A_p \ =\ \left\langle \xi^p, y \right\rangle 
\Big/ {\textstyle \sum_{i = 1}^{N} \xi_i^p}~.
\label{A_p}
\end{equation}
Note that the $\xi_i^p$ are positive, representing the
neural firing rate in attractor states. 

%%%%%%%%%%%%%%%%%%%%%%%%%%%%%%%%%%%%%%%%%%%%%
\begin{figure}[t]
\centering
\includegraphics[width=0.85\columnwidth,angle=0]{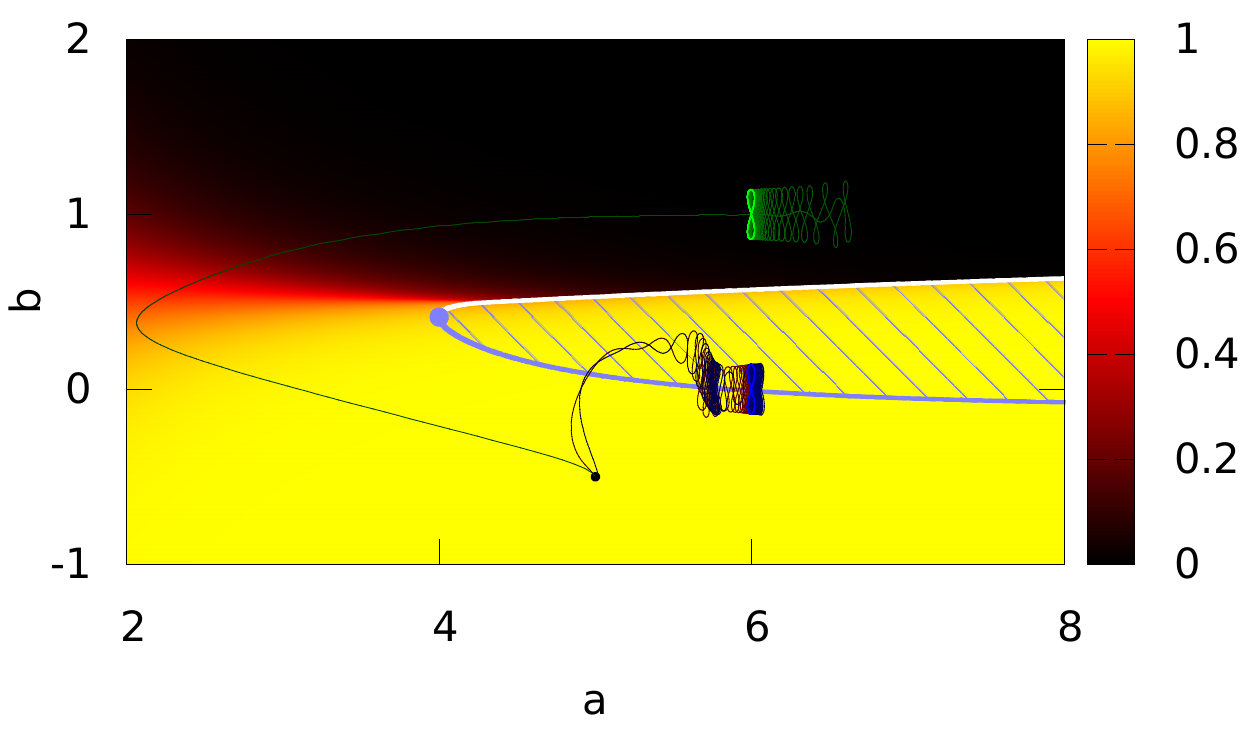}
\vspace{-1cm}
\caption{The trajectories of $(a_i(t),b_i(t))$ for finite
adaption rates, $\epsilon_a=0.1$ and $\epsilon_b=0.01$,
superimposed onto the phase diagram of the three-site network
shown in Fig.~\ref{fig:3-site_phaseDiagram}. The color encodes
the activity of the most active site. Inside the shaded blue area
two attractors are stable, outside there is only a single global 
attractor. The parameters are the same as for the simulation presented 
in Fig.~\ref{fig:three_site_neurons_time_series}. The thresholds
$b_1(t)$ and $b_3(t)$ (red/blue trajectories) oscillate across 
the second-order phase boundary (blue line), the threshold 
$b_2(t)$ (green trajectory) acquires a large value, receiving 
two excitatory inputs, and avoids the first-order phase 
boundary (white line) during the initial transient. All three 
trajectories start in the center, lower-half of the 
phase diagram, at $a=5$, $b=-0.5$ (black dot). 
Note that the underlying phase diagram is for 
identical thresholds $b_i$ and gains $a_i$.
}
\label{fig:adaption_in_phase_diagram}
\end{figure}
%%%%%%%%%%%%%%%%%%%%%%%%%%%%%%%%%%%%%%%%%%%%%

The data for $O_p$ and $A_p$ presented in
Fig.~\ref{fig:three_site_neurons_time_series} shows 
that the attractors become unstable in the presence of
finite polyhomeostatic adaption, retaining however a prominent
role in phase space. Unstable attractors, transiently 
attracting the phase space flow, can be considered to act 
as `attractor relicts'. The resulting dynamical system 
is hence an attractor relict network, viz a network of 
coupled attractor ruins \cite{gros2007,gros09}.
When coupling an attractor network to slow intrinsic
variables, here the threshold $b$ and the gain $a$, the
attractors are destroyed but retain presence in the flow
in terms of attractor ruins, with the dynamics slowing
down close to the attractor relict. 

The concept of an attractor ruin is quite general
and attractor relict networks have implicitly been generated
in the past using a range of functionally equivalent schemes.
It is possible to introduce additional local slow variables, 
like a reservoir, which are slowly depleted when a unit is 
active \cite{gros2007,gros09}. Similarly, slowly accumulating 
spin variables have been coupled to local dynamic thresholds in 
order to destabilize attractors \cite{horn1989}, and
local, slowly adapting, zero state fields have been considered
in the context of latching Potts attractor networks 
\cite{russo2008,akrami2012,treves2005,kropff2007}. 
Alternatively, attractors have been destabilized in neural
networks by introducing slowly adapting asymmetric components 
to the inter-neural synaptic weights \cite{sompolinsky1986}.

%%%%%%%%%%%%%%%%%%%%%%%%%%%%%%%%%%%%%%%%%%%%%
\begin{figure}[t]
\centering
\includegraphics[width=0.7\columnwidth,angle=0]{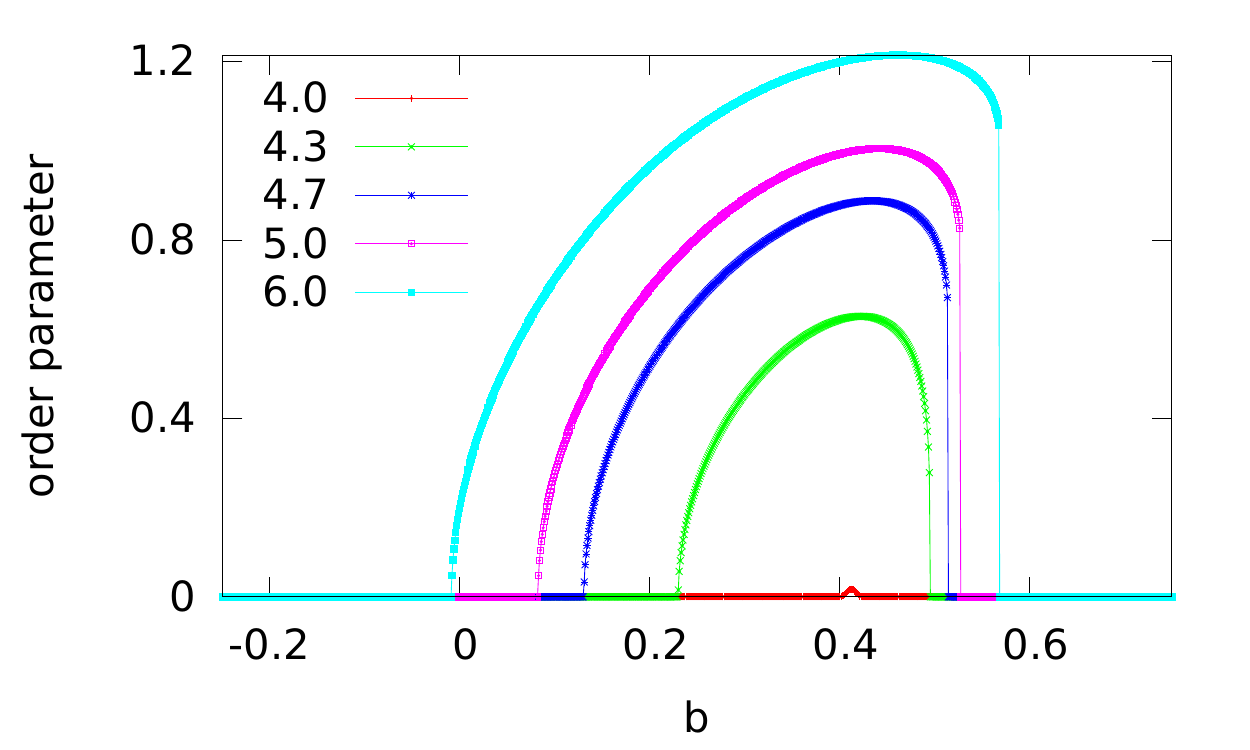}
\caption{Vertical cuts (for constant gains
$a=4,0,\ 4.3,\ 4.7,\ 5.0,\ 6.0$, bottom to top curves) 
through the phase diagram of the 3-site graph 
shown in Fig.~\ref{fig:3-site_phaseDiagram}.
Plotted is the order parameter
$\sqrt{\sum_{i=1}^3 \left(y_i^{(1)}-y_i^{(2)}\right)^2}$, where
$y^{(\gamma)}=(y_1^{(\gamma)},y_2^{(\gamma)},y_3^{(\gamma)})$
are the two fixpoint solutions ($\gamma=1,2$) in the 
coexistent region. The lower/upper transitions are
second/first order.
}
\label{fig:order-paramter}
\end{figure}
%%%%%%%%%%%%%%%%%%%%%%%%%%%%%%%%%%%%%%%%%%%%%

%--------------------------------------------
\subsection{Phase boundary adaption}
\label{subsec:phase-baundary}
%--------------------------------------------

In Fig.~\ref{fig:adaption_in_phase_diagram} we have 
superimposed, for the three-site network, a typical set of
trajectories, with finite adaption rates $\epsilon_a=0.1$ 
and $\epsilon_b=0.01$, onto the phase diagram evaluated
for non-adapting neurons, with $\epsilon_a=0=\epsilon_b$,
as shown in Fig.~\ref{fig:3-site_phaseDiagram}. The 
three trajectories 
(green: $(a_2,b_2)$, red and blue: $(a_1,b_1)$ 
and $(a_3,b_3)$)
all start well in the region with a single global
fixpoint, at $a_i=5$, $b_i=-0.5$ ($i=1,2,3$). Two
features of the flow are eye-catching.
\begin{itemize}
\item The intrinsic parameters of the active neurons
      settle, after a transient initial period,
      at the phase boundary and oscillate
      alternating between the phase with a single stable
      fixpoint and the phase with two attractors. 
\item The trajectory $(a_2,b_2)$ for the central neuron
      adapts to a large value for the threshold, taking care
      not to cross any phase boundary during the initial
      transient trajectory.
\end{itemize}
We first discuss the physics behind the second 
phenomenon. The Euclidean distance in phase space 
between the two attractors, when existing, can be 
used as an order parameter. In Fig.~\ref{fig:order-paramter} 
the order parameter is given for vertical cuts through 
the phase diagram. The lower transition is evidently
of second order, which can be confirmed by a stability
analysis of (\ref{three_sites}), with the upper transition 
being of first order, in agreement with a graphical 
analysis of (\ref{three_sites}). A first order transition 
will not be crossed, in general, by an adaptive process, 
as there are no gradients for the adaption to follow. 
This is the reason that the trajectory for $(a_2,b_2)$ 
shown in Fig.~\ref{fig:adaption_in_phase_diagram} 
needs to take a large detour in order to arrive to its 
target spot in phase space. This behavior is independent 
of the choice of initial conditions.

Using an elementary stability analysis one can show that
the condition for the second-order line is
$1=a\tilde y(1-\tilde y)$, where $\tilde y\equiv y_1=y_3$
(for the single global fixpoint). Only a single global attractor is 
consequently stable for $a\le4$ (since $\tilde y\in[0,1]$ 
and $\tilde y(1-\tilde y)\le0.25$) and the second 
and the first order lines meet at $(a_c,b_c)=(4,0.413)$, 
where the critcal threshold $b_c$ is determined by the 
self-consistent solution of $b_c=1/[1+\exp(4b_c-4)]-1/2$.
The trajectories $(a_1,b_1)$ and $(a_3,b_3)$ are hence 
oscillating across the locus in phase space where there 
would be a second-order phase transition, compare
Fig.~\ref{fig:adaption_in_phase_diagram}, for the
case of identical internal parameters $a_i$ and $b_i$.
With adaption, however, the respective internal parameters
take distinct values, $(a,b\pm\delta b)$ for the first/third
neuron and $(a,b_2)$ for the second neuron, with 
$a\approx 6$, $b\approx 0$ and $b_2\approx 1$. In fact
the system adapts the intrinsic parameters dynamical 
in a way that a non-stopping sequence of states
\begin{equation}
\dots \ \rightarrow\
\xi^1 \ \rightarrow\ \xi^0 \ \rightarrow\
\xi^2 \ \rightarrow\ \xi^0 \ \rightarrow\
\xi^1 \ \rightarrow\ \dots
\label{sequenceStates}
\end{equation}
is visited, where $\xi^1=(1,1,0)$, $\xi^2=(0,1,1)$ denote 
the two stable cliques and $\xi^0=(1,1,1)$ the fixpoint in 
the phase having only a single global attractor.
It is hence not a coincidence that the system adapts
autonomously, as shown in Fig.~\ref{fig:adaption_in_phase_diagram},
to a region in phase space where there would be a second
order phase transition for identical internal parameters.
At this point, small adaptive variations make the sequence
(\ref{sequenceStates}) easily achievable. The adaption
process hence shares some aspects with what one calls
'self-organized quasi-criticality' in the context of 
absorbing phase transitions \cite{SOCreview2013}.

%%%%%%%%%%%%%%%%%%%%%%%%%%%%%%%%%%%%%%%%%%%%%
\begin{figure}[t]
\centering
\includegraphics[height=0.23\columnwidth]{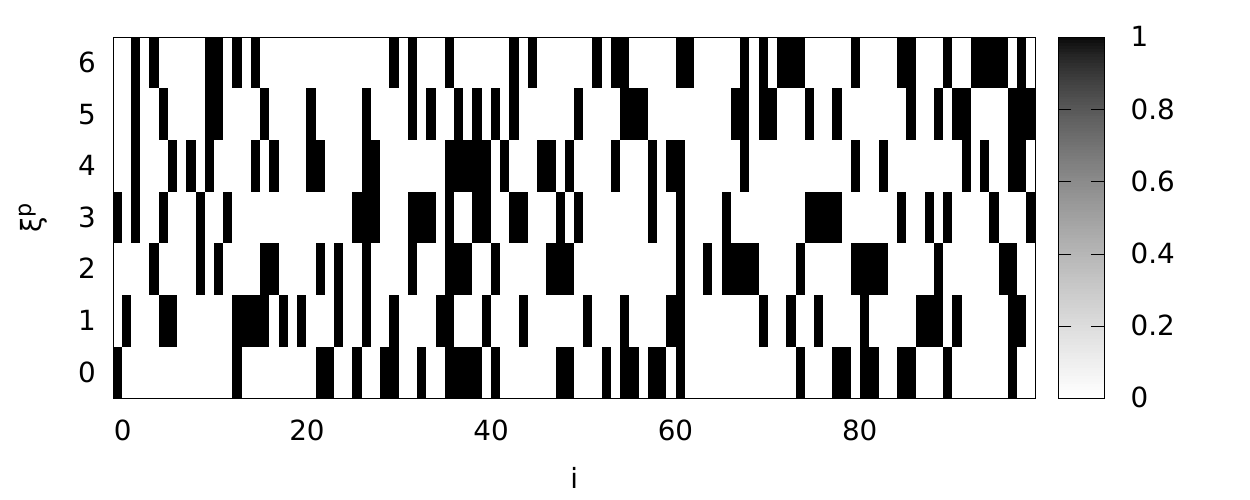}
\includegraphics[height=0.23\columnwidth]{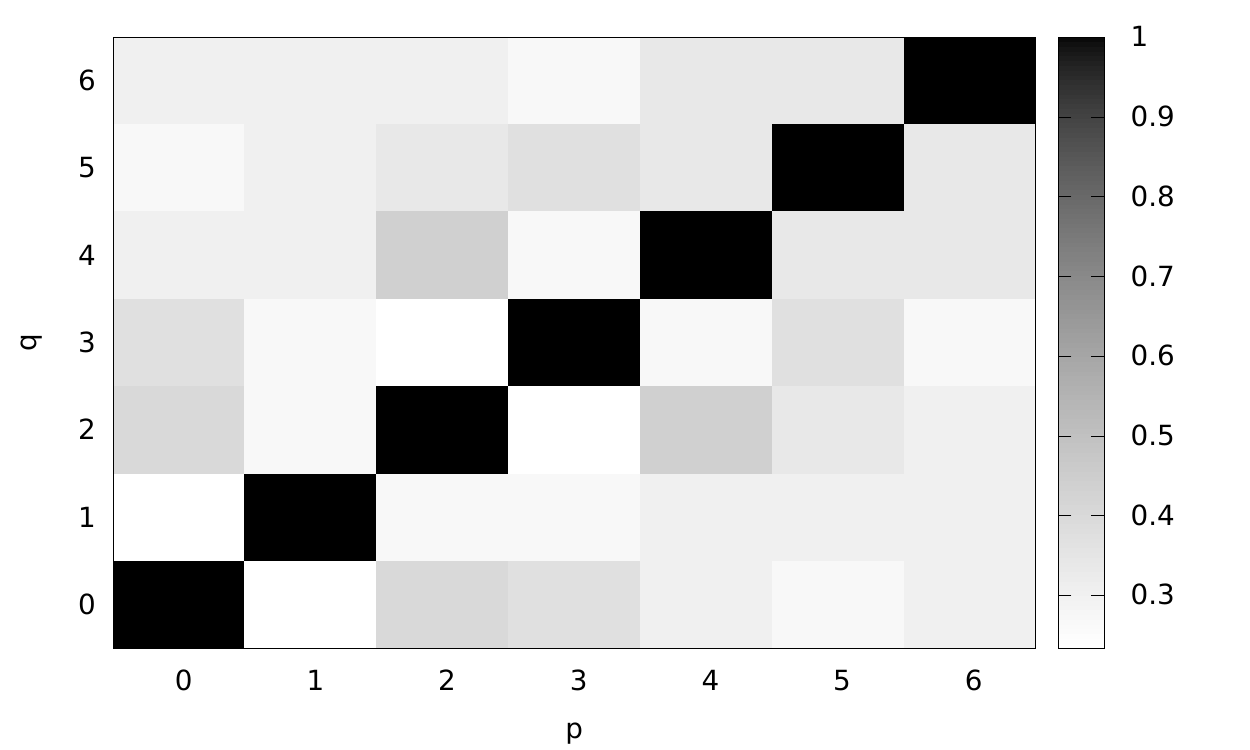}
\caption{Left: Example of $N_P=7$ random binary patterns $\xi^p$,
compare (\ref{xi_p_binary}), with sparseness~$\alpha = 0.3$ 
for a network of $N=100$ neurons. Right:
Respective pairwise mutual pattern overlaps~$O(\xi^{p}, \xi^{q})$.
Patterns maximally overlap with themselves, the inter-pattern
overlap is random and small.}
\label{fig:examplePatterns}
\end{figure}
%%%%%%%%%%%%%%%%%%%%%%%%%%%%%%%%%%%%%%%%%%%%%

%%%%%%%%%%%%%%%%%%%%%%%%%%%%%%%%%%%%%%%%%%%%%%%%%%%%%%%%%%%5
%%%%%%%%%%%%%%%%%%%%%%%%%%%%%%%%%%%%%%%%%%%%%%%%%%%%%%%%%%%5
\section{Latching in Hopfield networks with objective function stress}

We will now investigate networks of larger size $N$.
In principle we could generate random synaptic
link matrices $\{w_{ij}\}$ and find their eigenstates
$\xi^p=(\xi_i^p,\dots,\xi_N^p)$ using a diagonalization
routine. Alternatively we can use the Hopfield encoding
\cite{hopfield1982,hopfield1984} 
\begin{equation}
w_{i j} \propto \frac{1}{\alpha (N-1)} 
\sum_{p = 1}^{N_P} \left( \xi_i^p - \overline{\xi_i} \right) 
\left( \xi_j^p - \overline{\xi_j} \right),
\qquad\quad
\overline{\xi_i} = \frac{1}{N_P}
\sum_{p=1}^{N_P} \xi_{i}^{p}
\label{hopfield_encoding}
\end{equation} 
for the synaptic weights $w_{i j}$, where $\overline{\xi_i}$ 
are the arithmetic means of all pattern activities, for the 
respective sites.  Here $N_p$ is the number of encoded 
attractors and $\alpha\in[0,1]$ the mean pattern activity,
\begin{equation}
\alpha\ =\ \frac{1}{N}\sum_i x_i^p~.
\label{def_alpha}
\end{equation}
When using the Hopfield encoding (\ref{hopfield_encoding}),
the attractors are known to correspond, to a close
degree, to the stored patterns $\xi^p$, as long as
the number $N_p$ of patterns is not too large
\cite{hopfield1982,hopfield1984}. We here make use
of the Hopfield encoding for convenience, no claim
is made that memories in the brain are actually 
stored and defined by (\ref{hopfield_encoding}).

We consider in the following random binary patterns, as 
illustrated in Fig.~\ref{fig:examplePatterns}, 
\begin{equation}
\xi_i^p = \left\{
\begin{array}{cl} 
1 & \mathrm{with\ probability}\ \  \alpha \\ 
0 & \mathrm{with\ probability}\ \  1-\alpha 
\end{array} \right.
\label{xi_p_binary}
\end{equation}
where $\alpha$ is the mean activity level or sparseness.
The patterns $\xi^p$ have in general a finite, albeit small,
overlap, as illustrated in Fig.~\ref{fig:examplePatterns}.

%-------------------------------------------
%-------------------------------------------
\subsection{Regular latching dynamics in the absence of stress}

The target distribution $q(y)$ for the intrinsic
adaption has an expected mean
\begin{equation}
\mu \ = \ \int_0^1 \mathrm{d}y\,y\,q(y) 
\ =\Big|_{\lambda_2=0} \ \ 1 - \frac{1}{\lambda_1} + 
\frac{1}{e^{\lambda_1} - 1}~,
\label{eqn_lambda}
\end{equation}
which can be evaluated noting that the support of
$q(y)$ is $[0,1]$. The target mean activity $\mu$ can
now differ from the average activity of an attractor
relict, the mean pattern activity $\alpha$, compare
(\ref{def_alpha}). The difference between
the two quantities, viz between the two objectives,
energy minimization vs.\ polyhomeostatic optimization,
induces stress into the latching dynamics, which we
will study in the following.

%%%%%%%%%%%%%%%%%%%%%%%%%%%%%%%%%%%%%%%%%%%%%
\begin{figure}[t]
\centering
\includegraphics[width=0.8\columnwidth]{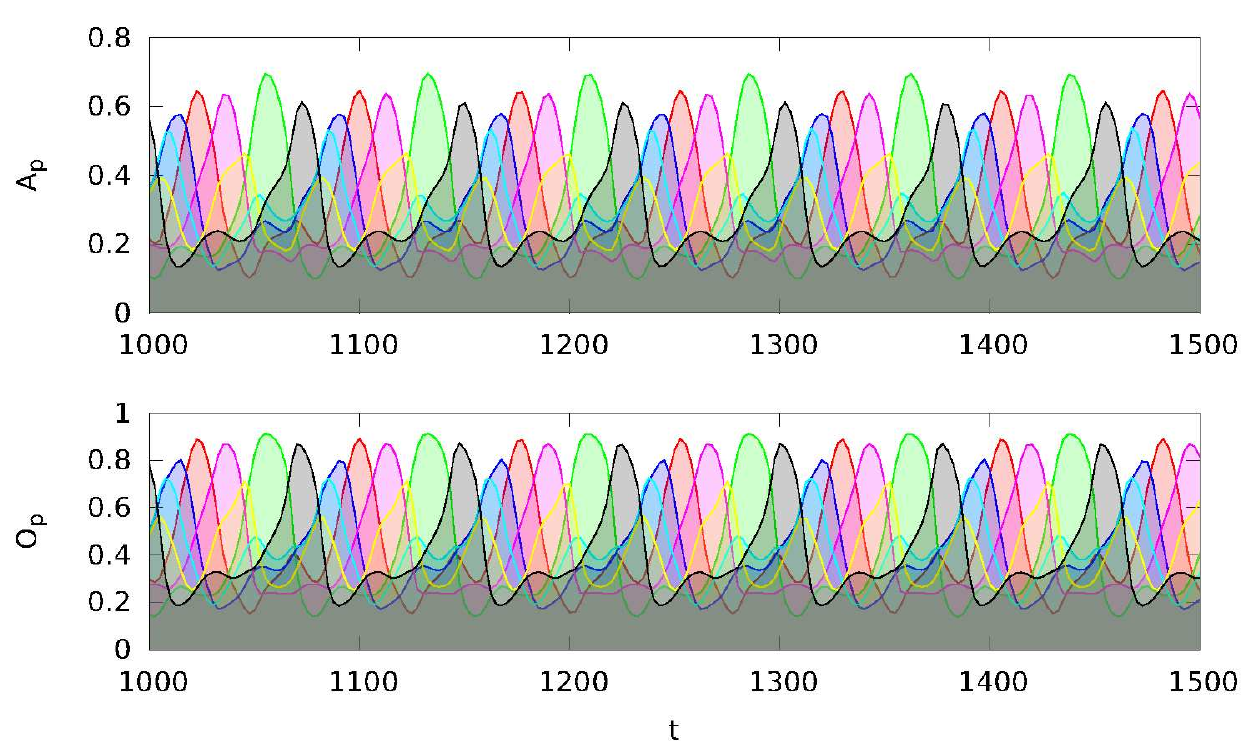}
\caption{Overlaps $O_p$ and $A_p$ (color encoded,
see Eqs.~(\ref{O_p}) and (\ref{A_p})), for a $N=100$ network,
of the neural activities with the $N_p=7$ attractor ruins, 
as a function of time $t$. The sparseness of the binary patterns 
defining the attractor ruins is $\alpha = 0.3$. Here the average 
activity $\alpha$ of the attractor relicts and the target mean 
activity level $\mu=0.3$ have been selected to be equal, resulting 
in a limiting cycle with clean latching dynamics.
}
\label{fig:patternLatchingTimeSeries_normal}
\end{figure}
%%%%%%%%%%%%%%%%%%%%%%%%%%%%%%%%%%%%%%%%%%%%%

\begin{figure}[t]
\centering
\includegraphics[width=0.8\columnwidth]{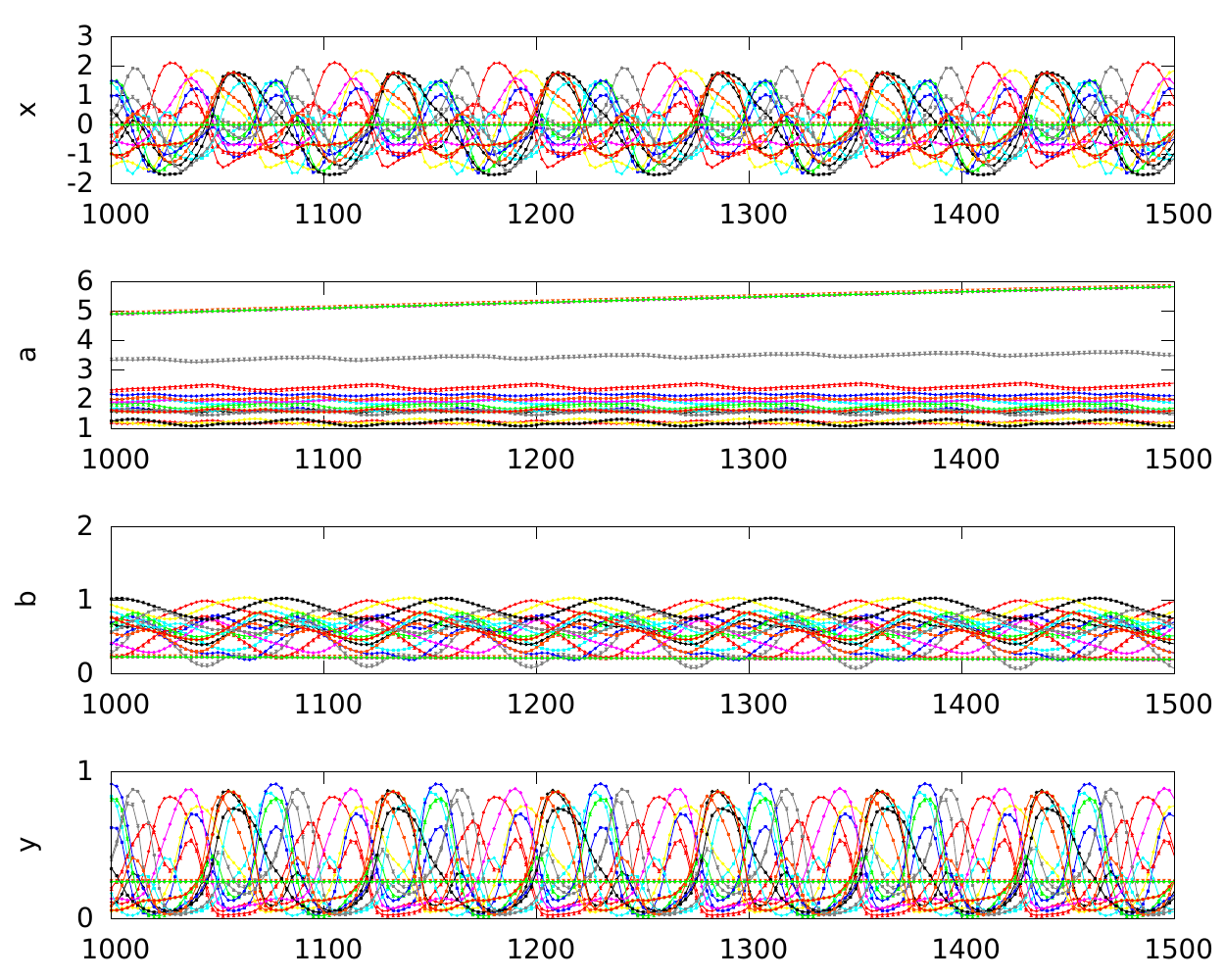}
\caption{Time evolution, for a selection of 20 out of $N=100$ neurons,
of the membrane potential $x_i$, gain $a_i$, threshold $b_i$ and firing 
rate $y_i$ for the time series of overlaps presented in 
Fig.~\ref{fig:patternLatchingTimeSeries_normal}.
} 
\label{fig:neuronTimeSeries_normal}
\end{figure}

We simulated Eqs.~(\ref{xdot}) and (\ref{abdot}) using 4th order
classical Runge-Kutta \cite{press2007} and a timestep of
$\Delta t = 0.1$. The resulting dynamics is dependent on
the magnitude of the adaption rates for the gain $a$ 
and for the threshold $b$. In general latching dynamics
is favored for small $\epsilon_b \sim 0.01$ and somewhat
larger $\epsilon_a \sim 0.01 \dots 1$. We kept a constant
leak rate $\Gamma = 1$. The results presented in
Figs.~\ref{fig:patternLatchingTimeSeries_normal} and
\ref{fig:neuronTimeSeries_normal} have been obtained
setting $\lambda_2 = 0$ and using
Table~1 for $\lambda_1$

%%%%%%%%%%%%%%%%%%%%%%%%%%%%%%%%%%%%%%%%%%%%%%
\begin{table}[b]
\label{tab_lambda}
\caption{Relation between the parameter~$\lambda_1$, for $\lambda_2 = 0$,
and the mean value~$\mu$, as given by Eq.~\ref{eqn_lambda},
for the target distribution $q(y)$.}
\hfill \begin{tabular}{c|ccccccccc}
$\mu$ & 0.1 & 0.2 & 0.3 & 0.4 & 0.5 & 0.6 & 0.7 & 0.8 & 0.9 \tabularnewline
\hline
$\lambda_1$ & -9.995 & -4.801 & -2.672 & -1.229 & 0 & 1.229 & 2.672 & 4.801 & 9.995 \tabularnewline
\end{tabular}
\end{table}
%%%%%%%%%%%%%%%%%%%%%%%%%%%%%%%%%%%%%%%%%%%%%%

In Fig.~\ref{fig:patternLatchingTimeSeries_normal} 
we present the time evolution of the overlaps
$A_p$ and $O_p$ for $\alpha=0.3=\mu$ and adaption
rates $\epsilon_a=0.1$, $\epsilon_b=0.01$. In this
case the target mean activity, $\mu$, of the
intrinsic adaption rules (\ref{abdot}) is consistent
with the mean activity $\alpha$ of the stored patterns
$\xi^p$, viz with the mean activity level of the 
attractor relicts. One observes that the system
settles into a limiting cycle with all seven stored 
patterns becoming successively transiently active,
a near-to-perfect latching dynamics. The dynamics is very
stable and independent of initial conditions, which were
selected randomly.

%%%%%%%%%%%%%%%%%%%%%%%%%%%%%%%%%%%%%%%%%%%%%
%%%%%%%%%%%%%%%%%%%%%%%%%%%%%%%%%%%%%%%%%%%%%

In Fig.~\ref{fig:neuronTimeSeries_normal} we present
the time evolution, for 20 out of the $N=100$ neurons,
of the respective individual variables. The simulation parameters
are identical for Figs.~\ref{fig:patternLatchingTimeSeries_normal}
and \ref{fig:neuronTimeSeries_normal}.
Shown in Fig.~\ref{fig:neuronTimeSeries_normal} are
the individual membrane potentials $x_i(t)$, the
firing rates $y_i(t)$, the gains $a_i(t)$ and the
thresholds $b_i(t)$. The latching activation of
the attractor relicts seen in
Fig.~\ref{fig:patternLatchingTimeSeries_normal}
reflects in corresponding transient activations 
of the respective membrane potentials and firing rates.
The oscillations in the thresholds $b_i(t)$ drive the
latching dynamics, interestingly, even though the
adaption rate is larger for the gain.

The synaptic weights $w_{ij}$ are symmetric and 
consequently also the overlap matrix presented 
in Fig.~\ref{fig:examplePatterns}. The latching
transitions evident in Fig.~\ref{fig:patternLatchingTimeSeries_normal}
are hence spontaneous in the sense that they are not
induced by asymmetries in the weight matrix
\cite{sompolinsky1986}. We have selected uncorrelated patterns
$\xi^p$ and the chronological order of the transient
states is hence determined by small stochastic differences
in the pattern overlaps. It would however be possible to consider 
correlated activity patters $\xi^p$ incorporating a rudimental 
grammatical structure \cite{treves2005,kropff2007}, which is
however beyond the scope of the present study.

%%%%%%%%%%%%%%%%%%%%%%%%%%%%%%%%%%%%%%%%%%%%%
\begin{figure}[t]
\centering
\includegraphics[width=0.8\columnwidth]{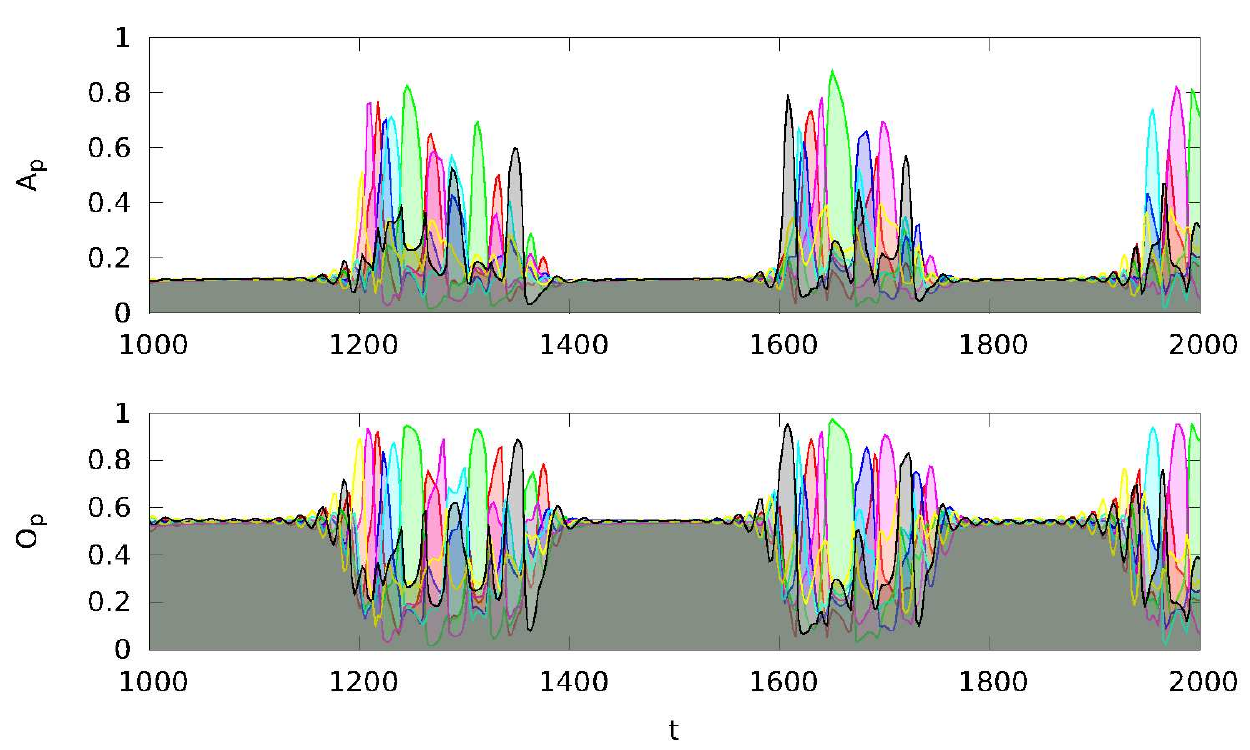}
\caption{Time evolution of overlaps $O_p$ and $A_p$ 
(color coded, compare Eqs.~(\ref{O_p}) and (\ref{A_p})) 
for all $N_P=7$ binary patterns, with sparseness~$\alpha = 0.3$, for a
network of $N=100$ neurons. The target mean neural activity is
$\mu = 0.15$, the difference between the two objective functions,
viz between $\mu$ and $\alpha$ induces stress. The intermittent
latching dynamics has a mean activity of about $0.3$,
which is too large. The phases of laminar flows between the
burst of latching has a reduced average activity, thus reducing
the time-averaged mean activity level toward the target $\mu=0.15$.  
} 
\label{fig:patternLatchingTimeSeries_bursting}
\end{figure}
%%%%%%%%%%%%%%%%%%%%%%%%%%%%%%%%%%%%%%%%%%%%%

%-------------------------------------------
%-------------------------------------------
\subsection{Stress-induced intermittent bursting latching}

In Fig.~\ref{fig:patternLatchingTimeSeries_bursting}
we present the time evolution of the overlaps $A_p$
and $O_p$ for the case when the two objective functions,
the energy functional and the polyhomeostatic optimization,
incorporate conflicting targets. We retain the average
sparseness $\alpha=0.3$ for the stored patterns, as for
Fig.~\ref{fig:patternLatchingTimeSeries_normal}, but 
reduced the target mean firing rate to $\mu=0.15$.
This discrepancy between $\alpha$ and $\mu$ induces
stress into the dynamics. The pure latching dynamics,
as previously observed in Fig.~\ref{fig:patternLatchingTimeSeries_normal},
corresponds to a mean activity of about $0.3$, in conflict
with the target value $\mu=0.15$.

Phase of laminar flow are induced by the objective function stress,
and the latching dynamics occurs now in the form of intermittent
bursts. The neural activity, see $A_p$ in 
Fig.~\ref{fig:patternLatchingTimeSeries_bursting}, is substantially
reduced during the laminar flow and the time averaged mean
firing rate such reduced towards the target activity level of
$\mu=0.15$. The trajectory does not come close to any particular
attractor relict during the laminar flow, the overall
$O_p$ being close to $0.5$ for all $N_p=7$ stored pattern.

The time evolution is hence segmented into two distinct phases,
a slow laminar phase far from any attractor and intermittent
phases of bursting activity in which the trajectories linger 
transiently close to attractor relicts with comparatively fast 
(relative to the time scale of the laminar phase)
latching transitions. Intermittent bursting dynamics has
been observed previously in polyhomeostatically adapting
neural networks with randomly selected synaptic weights
\cite{markovic10,markovic12}, the underlying causes
had however not been clear. Using the concept of competing
generating functionals we find that objective function 
stress is the underlying force driving the system
into the intermittent bursting regime.

%-------------------------------------------
%-------------------------------------------
\subsection{Robustness of transient state dynamics}

\begin{figure}[t]
\centering
\includegraphics[width=0.8\columnwidth]{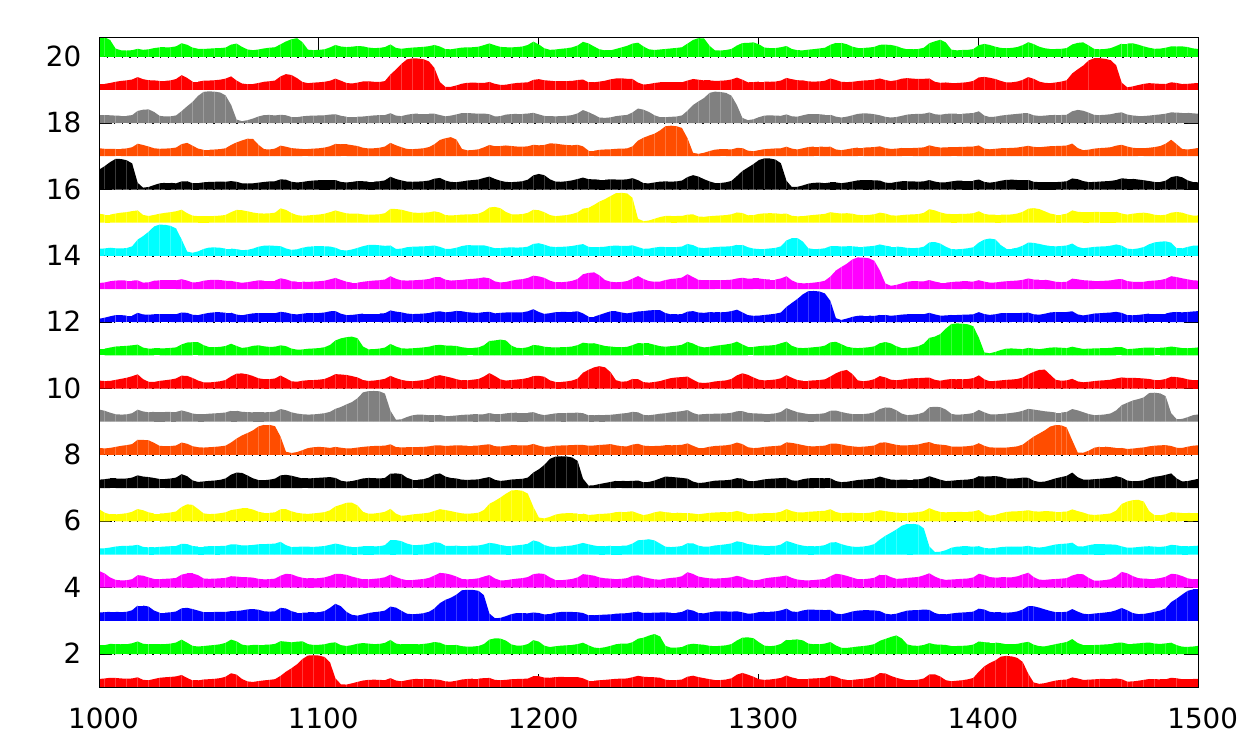}
\caption{Time evolution, for $N=1000$, of the overlaps $O_p$ for all
$N_P=20$ binary patterns (vertically displaced) with sparseness
$\alpha = 0.2$ and a target mean activity of
$\mu = 0.2$. There is no objective-function stress and the latching
is regular, the adaption rates are $\epsilon_a = 0.1$, $\epsilon_b = 0.01$.} 
\label{fig:patterns_latching_time_series_nudles}
\end{figure}

The latching dynamics presented in
Figs.~\ref{fig:patternLatchingTimeSeries_normal} and
\ref{fig:patternLatchingTimeSeries_bursting} is robust
with respect to system size $N$. We did run the simulation 
for different sizes of networks with up to $N=10^{5}$ neurons,
a series of sparseness parameters $\alpha$ and number
of stored patterns $N_P$. As an example we present in
Fig.~\ref{fig:patterns_latching_time_series_nudles}
the overlap $O_p$ for $N=1000$ neurons and
$N_P=20$ binary patterns with a sparseness of $\alpha=0.2$. 
No stress is present, the target mean activity level is $\mu=0.2$.
The latching dynamics is regular, no intermittent bursting
is observed. There is no constraint, generically, to force the
limiting cycle to incorporate all attractor relicts. Indeed,
for the transient state dynamics presented in
Fig.~\ref{fig:patterns_latching_time_series_nudles},
some of the stored patterns are never activated.

The data presented in 
Figs.~\ref{fig:patternLatchingTimeSeries_normal},
\ref{fig:patternLatchingTimeSeries_bursting} and
\ref{fig:patterns_latching_time_series_nudles}
is for small numbers of attractor relicts $N_p$,
relative to the systems size $N$, a systematic study
for large values of $N_p$ is beyond the scope of the
present study. Latching dynamics tends to break down,
generically speaking, when the overlap between distinct 
attractors, as shown in Fig.~\ref{fig:examplePatterns}, 
becomes large. The autonomous dynamics then becomes irregular.

%%%%%%%%%%%%%%%%%%%%%%%%%%%%%%%%%%%%%%%%%%%%%
%%%%%%%%%%%%%%%%%%%%%%%%%%%%%%%%%%%%%%%%%%%%%

%%%%%%%%%%%%%%%%%%%%%%%%%%%%%%%%%%%%%%%%%%%%%%%%%%%%%%%%%%%5
%%%%%%%%%%%%%%%%%%%%%%%%%%%%%%%%%%%%%%%%%%%%%%%%%%%%%%%%%%%5

\section{Discussion}

The use of generation functionals has a long tradition
in physics in general and in classical mechanics in particular.
Here we point out that using several multivariate objective 
functions may lead to novel dynamical behaviors and an 
improved understanding of complex systems in general. We propose
in particular to employ generating functionals which are 
multivariate in the sense that they are used to derive the 
equations of motion for distinct, non-overlapping subsets of
dynamical variables. In the present work we have studied a neural
network with fast primary variables $x_i(t)$, the membrane
potentials, and slow secondary variables $a_i(t)$ and
$b_i(t)$, characterizing the internal behavior of individual
neurons, here the gains $a_i$ and the thresholds $b_i$. The
time evolution of these sets of interdependent variables
is determined respectively by two generating functionals:
\begin{description}
\item[\sc energy functional] \hspace{1ex}\newline
   Minimizing an energy functional generates the equation of
   motion (leaky integrator) for the primary dynamical 
   variables $\dot x_i(t)$, the individual membrane potentials.
   
\item[\sc information theoretical functional] \hspace{1ex}\newline
   Minimizing the Kullback-Leibler divergence between the distribution
   $p_i(y)$ of the time-average neural firing rate $y_i$ and a
   target distribution function $q(y)$ maximizing information
   entropy generates intrinsic adaption rules $\dot a$ and
   $\dot b$ for the gain $a$ and the threshold $b$ 
   (polyhomeostatic optimization).
   
\end{description}
Generating functionals may incorporate certain targets or 
constraints, either explicitly or implicitly. We denote the 
interplay between distinct objectives incorporated by 
competing generating functionals {\em ``objective functions 
stress''}. For the two generating functionals considered in
this study there are two types of objective functions 
stress:
\begin{description}
\item[\sc functional stress] \hspace{1ex}\newline
  The minima of the energy functional are time-independent point 
  attractors leading to firing-rate distributions $p_i(y)$ which 
  are sharply peaked. The target firing-rated distribution $q(y)$ 
  for the information-theoretical functional is however smooth
  (polyhomeostasis). This functional stress leads to the formation 
  of an attractor relict network.
\item[\sc scalar stress] \hspace{1ex}\newline
  The mean target neural firing rate $\mu$ is a (scalar) 
  parameter for the target firing-rate distribution $q(y)$, 
  and hence encoded explicitly within the information theoretical 
  functional. The local minima of the energy functional, determined 
  by the synaptic weights $w_{ij}$, determine implicitly the mean 
  activity levels $\alpha$ of the corresponding point attractors. 
  Scalar objective function stress is present for $\alpha\ne\mu$.
\end{description}
For the two generating functionals considered, we find that the 
scalar objective function stress induces a novel dynamical
state, characterized by periods of slow laminar flow 
interseeded by bursts of rapid latching transitions.
We propose that objective function stress is a powerful tool, in
general, for controlling the behavior of complex dynamical systems. 
The interplay between distinct objective functions may hence serve 
as a mechanism for guiding self organization 
\cite{martius2007,prokopenko2009,haken2006}. 

% We acknowledge the support of the German Science Foundation.

%%%%%%%%%%%%%%%%%%%%%%%%%%%%%%%%%%%%%%%%%%%%%%%%%%%%%%%%%%%5
%%%%%%%%%%%%%%%%%%%%%%%%%%%%%%%%%%%%%%%%%%%%%%%%%%%%%%%%%%%5

\end{document}